# EMIC Wave Distributions Observed by the Van Allen Probes

Nezir Alic

August 2017

nalic@caltech.edu

# Contents

**Abstract**

*Note: This paper was originally submitted to the Siemens Competition for Math, Science, and Technology; therefore, all conventions used are those consistent with Siemens guidelines.*

Characteristics of electromagnetic ion cyclotron (EMIC) waves detected by the Van Allen probes are statistically analyzed, particularly wave-band, bandwidth, and time duration. The Electric and Magnetic Field Instrument Suite and Integrated Science, an instrument on board the Van Allen Probes, provides the necessary magnetic field measurements to examine 33 months of EMIC wave occurrence (1 September 2013 to 31 May 2016). Upon visual identification, the waves are grouped into their respective wave-bands, $H^+$, $He^+$, or $O^+$, defined by the frequencies they are observed at, manifested in the daily spectrograms as their location relative to gyrofrequency lines. Nearly 2,500 EMIC wave events are detected. Results suggest a prevalence of $He^+$-band waves, and a rarity of $O^+$-band waves (1,155 $H^+$-band events, 1,176 $He^+$-band events, and 125 $O^+$-band events). The most prevalent bandwidth range for events in general is found to be 0.25 - 0.5 Hz. However, this appears to vary among the three wave-bands. Helium and oxygen wave-band events tend to have shorter bandwidths (0.25 – 0.5 Hz) than their hydrogen counterparts. Time duration is more consistent among wave-bands, and while $H^+$-band events on average have a slightly shorter duration, the most common time duration for all wave-bands is between 20 and 40 minutes.

## 1. Introduction

Electromagnetic ion cyclotron (EMIC) waves are disturbances generated by energetic ions (10 -100 keV) [*Fraser et al., 2010*] in the Earth's magnetosphere. They are a component of the many factors determining space weather and driving its dynamic activities. Specifically, they are known to cause $He^+$ energization [*Zhang et al.,* 2010] and pitch-angle scattering loss [*Wang et* *al., 2014*]. The latter results in electron losses in the radiation belts and proton losses in the ring current [*Fraser et al., 2010*]. EMIC waves also heat cold electrons in the magnetosphere through Landau damping, thereby generating stable red auroral arcs [*Zhou et al., 2013*]. Increased knowledge of the nature of these oscillations could reveal their effects on the constitution of the Earth's atmosphere and would shed more light on the complex activities of the Van Allen radiation belts. Understanding them is thus also of central importance to spacecraft design, astronaut safety, and mission planning, all of which can be endangered by high-energy particles in the belts.

The bandwidth of EMIC waves ranges from 0.1 to 5 Hz (Pc1 – 1). The waves are classified into three wave-bands, depending on their location relative to gyrofrequency lines. Hydrogen ion ($H^+$) band wave events occur beneath the proton gyrofrequency and above the helium ion gyrofrequency. Similarly, helium ion ($He^+$) band exist under the helium ion gyrofrequency and above the oxygen ion gyrofrequency, and oxygen ion ($O^+$) band occur below the oxygen ion gyrofrequency.

EMIC waves are created by thermally anisotropic clouds of ions, often at intersections of the hot ring current and cold plasma in the plasmapause [*Criswell et al., 1969*]. They tend to occur in areas with large total plasma density and minimal magnetic field strength (i.e., the magnetic equator) [*Kennel and Petschek, 1966*]. Observations by CRRES have established their source region as ± ~11˚ Magnetic Latitude (MLAT) [*Loto'aniu et al., 2005*]. However, recently *Allen et al.* [2013] detected bidirectional wave packets at 33 - 49° MLAT, implying the possibility of wave excitation further from the magnetic equator. Many studies suggest the plasmapause and plasmaspheric plumes as particularly favorable region of generation for the waves [e.g., *Jordanova et al., 2007; Fraser et al., 1989; Horne and Thorne, 1993*]. $O^+$ wave-band events have been observed primarily within the plasmapause, closer to Earth than the other wave-bands [*AA. Saikin et al., 2015*]. In terms of distance from Earth, EMIC waves in general have been observed everywhere from L-shells L = 3 to L = 14. (An L-shell denotes the magnetic field line that crosses the magnetic equator at a distance L from Earth, measured in Earth radii $R_E$. As an example, L = 3 refers to the field line that crosses the equator 3 $R_E$ from Earth.)

Generation of EMIC waves may also be significantly intensified during magnetic storms, when hot ions are propelled into the inner magnetosphere [*Bräysy et al., 1998*]. Other studies conclude that this effect is more prevalent in the recovery phase of storms [*Bortnik et al., 2008*]. Certain time periods have also been cited as more conducive to EMIC wave generation. In general, EMIC wave activity is significantly greater on the dayside of the magnetosphere than on the nightside [*AA. Saikin et al., 2015*]. Peaks in $H^+$ and $He^+$ EMIC wave activity have been noticed in the afternoon magnetic local time (MLT) sector [*Morley et al., 2009; Pickett et al., 2010*]. A more recent study has identified a prenoon sector as well for this heightened activity, for all three wave-bands [*AA. Saikin et al., 2015*].

Since their discovery, EMIC waves have been the subject of various studies based on both ground and satellite observations [*Píša et al., 2015*]. These experiments have uncovered much new information, particularly regarding characteristics such as the waves' generation, normal angle, ellipticity, and most favorable regions and time periods of occurrence. Very few statistical studies, however [e.g., *Yu et al., 2015; AA. Saikin et al., 2015*], have been performed on $O^+$-band waves, largely due to previous instruments' inability to consistently observe them at the low frequencies at which they occur. This problem stemmed from inadequate resolution. The high orbit of the spacecraft typically used also presented difficulties ($O^+$-band events, as mentioned above, tend to lie in lower L-shells).

Moreover, many studies have been limited by their inability to cover all MLT. For example, the CRRES mission, although covering a relatively low L range (3 – 8), failed to complete a full precession and thus left a considerable deficiency of data between 8:00 and 14:00 MLTs [*AA. Saikin et al., 2015*]. Similarly, *Kasahara et al.* [1992], a study using the Akebono satellite, did not equally examine all MLTs, as Akebono only detected EMIC waves during equatorial crossings [*AA. Saikin et al., 2015*].

This study provides a thorough analysis of the inner magnetosphere (L = 2 – 8), analyzing an extensive sample of EMIC wave events in all MLT over 33 months of the Van Allen Probe mission. It examines all three wave-bands of EMIC waves, the inclusion of $O^+$-band events being made possible by the high-resolution and low orbit of the probes. The study adds to *Yu et al.* [2015] by considering data from both Van Allen probes (as opposed to solely probe A), and it adds to *AA. Saikin et al.* [2015] primarily by investigating the distributions of time duration and frequency range for the different wave-bands of wave events, an analysis that has not been carried out in previous papers.

## 2. Materials & Methods
### 2.1 Van Allen Probes

The Van Allen Probes, formerly known as the Radiation Belt Storm Probes [*Kessel et al., 2013*] are two spacecraft that orbit the Earth every 9 hours with a perigee of 1.1 $R_E$ and apogee of 5.8 $R_E$ [*Mauk et al., 2012*]. They were launched on 30 August 2012 in hopes of gaining a deeper understanding of the two Van Allen radiation belts that surround Earth [*Nasa*]. These belts, discovered in 1958, consist of charged particles, or plasma. As mentioned above, those particles can be dangerous to humans and their technologies, particularly during magnetic storms. The Van Allen probes, built largely to study changes in the belts, are identical in every respect except for their speed; one probe is faster and laps the other every 2.5 months [*AA. Saikin et al., 2015*]. They carry out a highly elliptical, low-inclination (~10˚) orbit [*Mauk et al., 2012*]. The probes complete one full precession every 22 months, and cover L-shells of L = 2 - 8 [*AA. Saikin et al., 2015*].

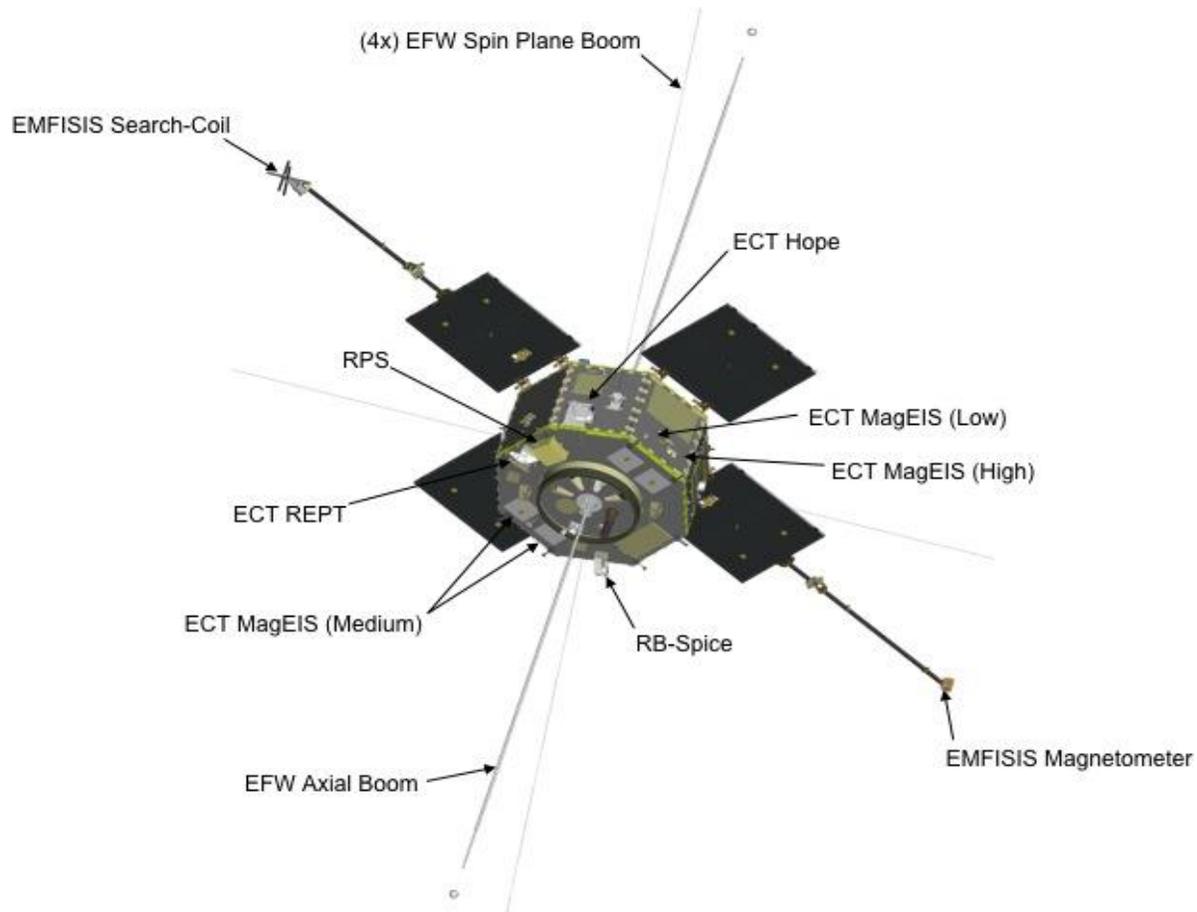

**Figure 1.** Components of a Van Allen probe are labelled above, including the EMFISIS Magnetometer, (bottom right) and search coil (top left) [*JHU*].

On board the probes is the Electric and Magnetic Field Instrument Suite and Integrated Science (EMFISIS), an instrument which measures the magnetic field with high temporal resolution (64 vectors per second) [*Kletzing et al., 2013*]. One of several recent magnetic field investigations created by researchers at NASA's Goddard Space Flight Center (GSFC), it consists of two sensors, a triaxial AC magnetic search coil magnetometer (MSC) and a triaxial fluxgate magnetometer (MAG) [*uiowa*]. A Main Electronics Box (MEB) stores and processes the signals received from these sensors [*Kletzing et al., 2013*]. Measurements from MAG, which come in the form of 3D magnetic field vectors [*Kletzing et al., 2013*], are utilized in this study [*AA. Saikin et al., 2015*].

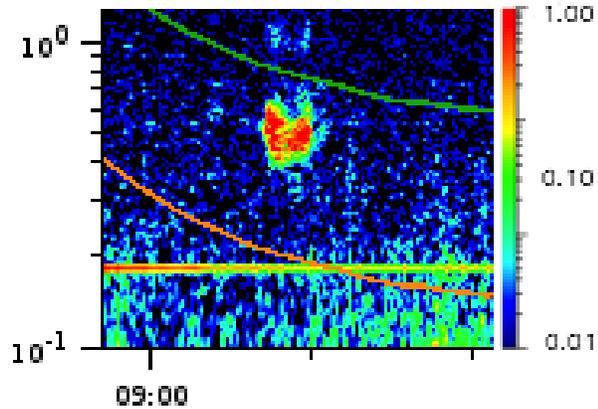

**Figure 2.** A sample $He^+$-band event, observed 2015 October 15. The horizontal axis quantifies time in hours and the vertical axis, on a log scale, quantifies frequency in Hertz. On the right is the bar denoting power ($nT^2$/Hz), also on a log scale. The green line represents the helium gyrofrequency line, and the orange represents the oxygen gyrofrequency line.

**2.2 Data Analysis**

Magnetic field data from the probes was converted via the fast Fourier transform (FFT) technique [*AA. Saikin et al., 2015*] into daily spectrograms, plots of time vs. frequency, which also displayed information such as MLT, L-shell, power, and gyrofrequency lines. These plots were given to me by a professor and I then examined them visually in order to identify EMIC wave events.

Events were required to meet several criteria in order to be classified as EMIC waves. Firstly, a minimum power of $0.1 nT^2$/Hz was established. To distinguish from background noise, each wave also must have remained visible for at least 5 minutes in universal time (UT). Lastly, an event was only flagged if it abided by the frequency range of 0.01 - 5 Hz. Each wave packet that satisfied these conditions was marked, a rectangle being drawn around it, and recorded on an excel spreadsheet that included the date on which it occurred, the probe it was observed by, its frequency range, start time, end time, and wave-band.

Events that crossed too many noise lines (manifestations of instrument noise) or seemed significantly contaminated by Ultra Low Frequency (ULF) waves were discounted. In cases where an EMIC wave spanned over multiple wave-bands (i.e., crossed a gyrofrequency line), its wave-band would be recorded as the one in which the majority of the wave packet occurred. Events that seemed to be captured twice, once by each probe, were documented as two distinct wave events.

## 3. Results

### 3.1 Band Distributions

In the 33 months (1.5 precessions) of the Van Allen Probes mission that were examined in this study, 2,456 EMIC wave events were observed. Table 1 shows the number of events by wave-band and by probe. Probe A observed 238 more events than Probe B, likely as a result of several brief gaps in coverage. $He^+$-band events comprise the greatest number of those detected (1,176, or 47.88%), and $H^+$-band events constitute an almost equivalent proportion (1,156 events, or 47.07%). $O^+$-band events, however, account for only a small minority (124 events, or 5.05%) of the recorded waves, a finding consistent with previous studies.

| Table 1. Wave Occurrence by Probe and Wave-Band | | | | |
|---|---|---|---|---|
|  | $H^+$ | $He^+$ | $O^+$ | Total |
| Probe A | 586 | 676 | 85 | 1347 |
| Probe B | 570 | 500 | 39 | 1109 |
| Total | 1156 | 1176 | 124 | 2456 |

| Table 2. EMIC Wave Occurrence by Wave-Band and Time | | |
|---|---|---|
| Wave-band | Number of Events | Total Time Duration (Hours) |
| $H^+$ | 1156 | 667:367 |
| $He^+$ | 1176 | 716:067 |
| $O^+$ | 124 | 68:65 |

The total time duration of each wave-band, displayed in table 2, was calculated by taking the difference between start and end times for each event, and summing the results. As expected, $He^+$-band events account for the greatest amount of wave observation time (716 hours, 4 minutes), $H^+$-band the second greatest amount (667 hours, 22 minutes), and $O^+$-band the least (68 hours, 39 minutes). Notably, the difference between the total time durations of Helium and Hydrogen wave-band events is considerably greater than the difference in their number of events, or percentage. In other words, a ~50-hour time difference (667 – 716) does not equate to a mere 20 events (1,176 – 1,156), or 0.71% (47.88 – 47.07) of the observed waves. Thus, Helium wave-band events likely tend to last longer than EMIC waves of other bands.

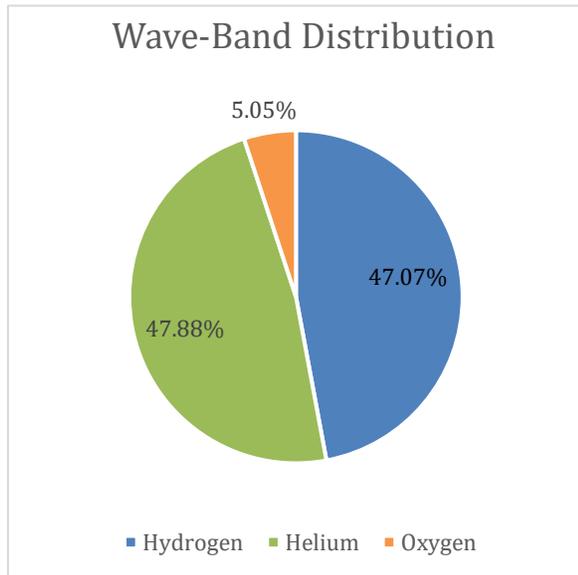

**Figure 3.** The distribution of observed EMIC waves by wave-band, given in percentages.

At left, figure 3 illustrates the percentage distribution of the different wave-bands. Here, the similar occurrence rate of $He^+$-band events and $H^+$-band events, as well as the relative rarity of $O^+$-band events, is made evident.

**3.2 Bandwidth Distributions**

Histograms were created to further examine the frequency with which the observed EMIC waves exhibited various bandwidths. To calculate the total bandwidth of all the recorded events, the difference between highest and lowest frequency of each event was obtained, and these differences were summed. The same process was repeated for each wave-band. Bins of 0.25 Hz were used after testing other less efficient bin sizes. Thus, an event under the "0.5" category, for example, has a bandwidth between 0.26 and 0.5, inclusive, while events labelled as "2" Hz have a bandwidth between 1.76 and 2, inclusive. Intervals that optimized the amount of information shown were also used for the vertical axes.

Overall, the most common range of bandwidth for EMIC waves detected in the inner magnetosphere during the 33-month period was found to be between 0.26 and 0.5 Hz (inclusive).

Approximately 750 waves were found in this range. The second most common was the 0.25 bin, with ~590 waves, followed by the 0.75 bin, with ~390 events. The number of observed EMIC waves decreased steadily for each bandwidth bin beyond 0.75 Hz. There were very few events (~30) detected with a bandwidth greater than 3 Hz. The vast majority exhibited a bandwidth less than or equal to 1 Hz.

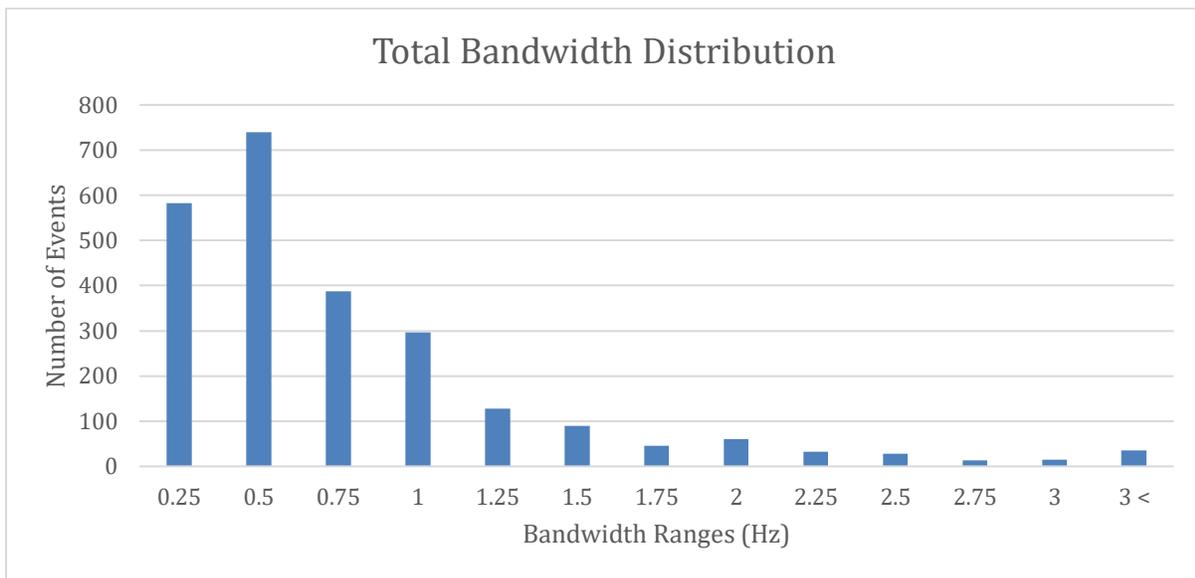

**Figure 4.** Histogram for the bandwidth distribution of all EMIC waves observed by the Van Allen probes from 1 September 2013 to 31 May 2016. Each bin's label defines a range of bandwidths ending at that number and beginning at the number that is 0.25 less than it (e.g., "0.5" encompasses 0.26 – 0.5). Note that the first bin represents only 0.1 to 0.25; this is because of the bandwidth criterion for EMIC wave events mentioned in the Methods section.

When broken up by wave-band, new trends appear. While $H^+$-band events maintain 0.5 as the most common bandwidth bin, $He^+$-band and $O^+$-band wave events exhibit a peak bandwidth of 0.25 Hz. $H^+$-band events only rarely demonstrate such a low bandwidth; less than 50 $H^+$-band EMIC waves with a bandwidth less than or equal to 0.25 were observed (of the total 1,156 $H^+$ events). The second most common bin for $H^+$-band waves was 0.75 Hz, followed by 1 Hz. The number of events with higher frequencies drops off sharply after this point, with only

~100 events being recorded in the 1.25 Hz bin. Very few waves of this band were found to have bandwidths greater than 3.5 (~20).

He$^+$ and O$^+$ band waves demonstrated such a decline in occurrence in lower bins. Followed by the two most populated bins, 0.25 and 0.5 Hz, respectively, He$^+$ band waves already drop off at 0.75 Hz, and very few exist with a bandwidth greater than 2.25 Hz. A dip in frequency of O$^+$-band events is also seen in the 0.75 bin. Even the 0.5 bin lacks a significant proportion of O$^+$-band waves (~25%). A considerable majority (81 of 124, or ~65%) O$^+$-band events possess a bandwidth less than or equal to 0.25 Hz.

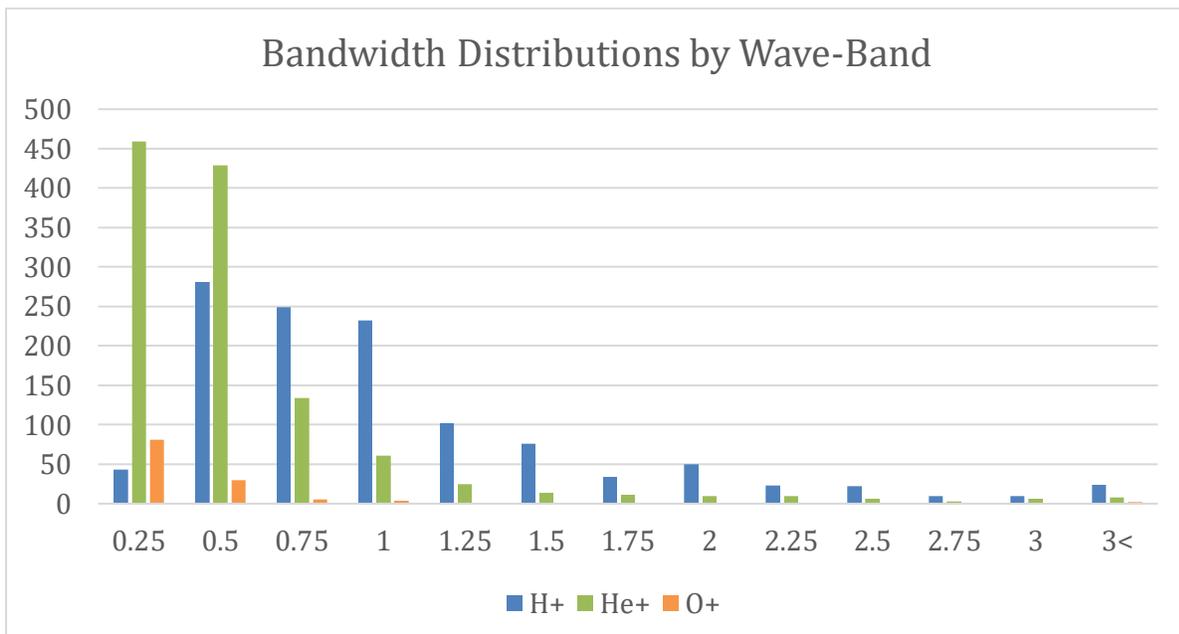

**Figure 5.** A histogram in the same format as Figure 4, showing bandwidth distribution by band. X-axis values represent bandwidth in Hertz, while y-axis values represent the number of events.

The pattern exhibited by the helium and oxygen bars (namely, a consistent decrease in number of events as one increases bin values) would also appear among the hydrogen bars if they were shifted over to the left by one bar. The significance of this is that H$^+$-band events have greater bandwidths, on average, than the other two bands of EMIC waves.

It should also be noted that the 0.25 and 0.5 bins contain nearly equal numbers of events for helium, while there is a large disparity between these two bins for the other two bands. As mentioned above, the 0.5 bin contains many more $H^+$-band events than does the 0.25 bin, and the reverse is true of $O^+$ band events. As can be seen in the hydrogen bars, the 0.5, 0.75, and 1 bins all hold a relatively similar number of events (between 232 and 281).

### 3.3 Time Duration Distributions

Similar plots were created to examine the frequency with which the observed EMIC waves exhibited various time durations. To calculate the total time duration of all the events, the difference between start time and end time for each event was obtained, and these differences were summed. The same process was repeated for each wave-band. For these histograms, bins of 20 (20 minutes) were used. As an example, an event under the "40" category has a time duration between 21 and 40 minutes, inclusive. Vertical axes were, again, used as most convenient.

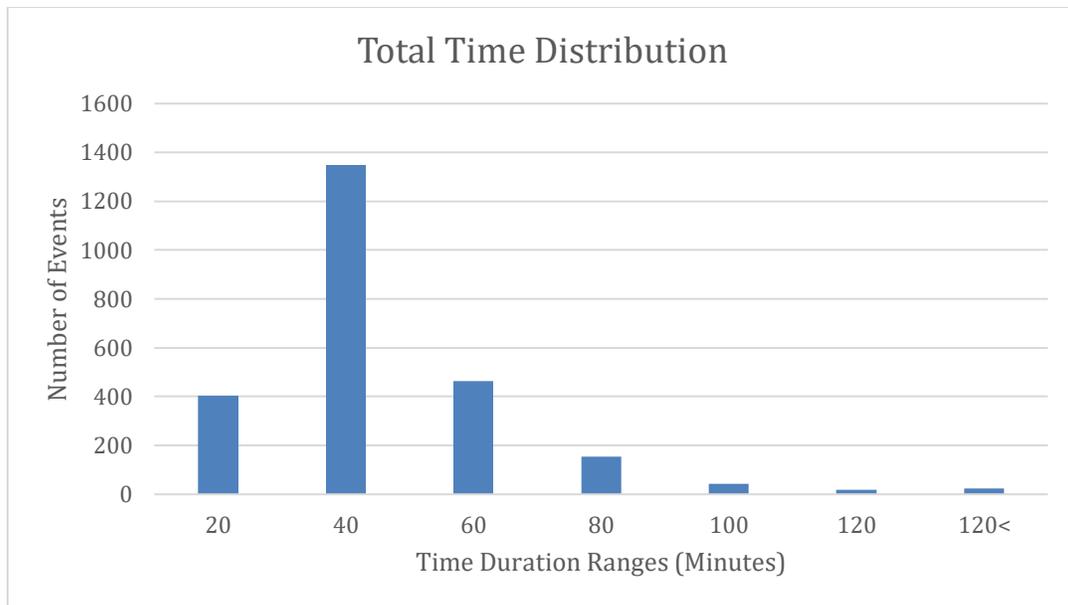

**Figure 6.** Histogram for the time duration distribution of all EMIC waves observed by the Van Allen probes from 1 September 2013 to 31 May 2016. Each bin's label defines a range of time durations ending at that value and beginning at the value 20 minutes less than it (e.g., "40" encompasses 21 – 40). Note that the first bin represents only 5 to 20; this is because of the time duration criterion for EMIC wave events mentioned in the Methods section.

For the ~2,500 waves detected, the most common range of time duration is 20 – 40 minutes. Approximately 1,350 events were found in this range. The second most prevalent is the 60 bin, with ~450 events, followed by the 20 bin, with ~400 events. The number of observed EMIC waves decreased steadily for each bandwidth bin beyond 60 minutes. Very few of the detected events (~25) lasted longer than 120 minutes. The vast majority lasted for less than or equal to 60 minutes.

Time durations across different wave-bands are more consistent than their frequency ranges. All three types of EMIC waves maintain the 40 minute bin as their peak time duration. For each wave-band, more than half of observed events lasted for a period of time within this 20 – 40 minute range. This is a very significant time duration preference. While the second most populated bin for $H^+$-band events (20) agrees with that of the general plot, that of $He^+$-band and $O^+$-band events does not. Those wave-bands appear to have a time duration of 60 minutes more often than 20 minutes. For each wave-band, there is a considerable drop in number of events for bins greater than 60 minutes. Very few events in any band last for longer than 100 minutes

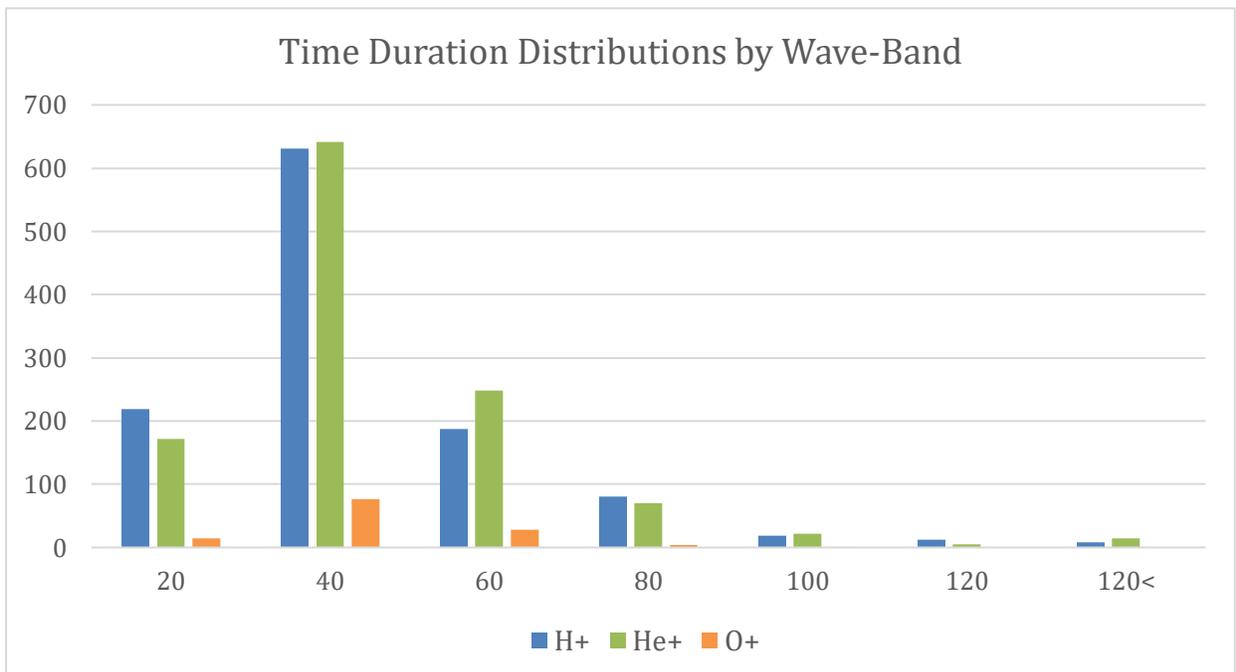

.

**Figure 7.** Histograms in the same format as Figure 6, showing time duration distribution by band. X-axis values represent time in minutes, while y-axis values represent the number of events.

This preference of $He^+$-band events for longer time durations reinforces the finding in table 2; $He^+$-band waves (as well as $O^+$-band) do appear to last longer, on average than their $H^+$-band counterparts.

## 4. Discussion

In this study, data from 33 months of the Van Allen Probes mission was examined visually, leading to the detection of nearly 2,500 $H^+$-, $He^+$-, and $O^+$–band EMIC wave events. Characteristics of these waves such as band, frequency, and time duration were statistically studied. This investigation serves as an expansion of previous EMIC wave studies [*Anderson et al., 1992a,1992b; Kasahara et al., 1992; Halford et al., 2010; Meredith et al., 2014; Allen et al., 2015; Yu et al., 2015; AA. Saikin et al., 2015*] primarily by observing all three wave-bands, covering all MLTs (as well as lower L-shells) consistently, using both Van Allen Probes, and specifically analyzing the distributions of bandwidth and time duration for the large number of waves observed.

The band distribution found in this paper largely agrees with that of others that have addressed the question. It is widely established that $O^+$-band waves are the least common of the three wave-bands [*Zhou et al., 2012*]; this paper reinforces this fact with its finding that only 5.05% of EMIC waves in the inner magnetosphere are $O^+$-band waves. However, two of the most recent assessments of band distribution do conflict in some notable ways with the distribution found in this paper. The first that will be discussed was performed by *AA. Saikin et al.* [2015]. It agrees with this paper's findings with regards to the order of the frequency of the bands – helium, then hydrogen, then oxygen band events. However, whereas this paper found a near equal amount of $He^+$-band and $H^+$-band waves (47.88% and 47.07%), the other determined a clear preference for $He^+$-band waves; (56.81% $He^+$-band waves, 34.37% $H^+$-band waves). It also obtained a slightly larger proportion of oxygen band events, 8.82% (compared with 5.05% in this paper).

The are several possible explanations for these discrepancies. Firstly, although *AA. Saikin et al.* [2015] also used the Van Allen probes and therefore surveyed the lower L-shells surveyed here, they studied a different time period than the one examined in this paper. Specifically, the 2015 study surveyed the 22 months from 8 September 2012 to 30 June 14, while this study covered 1 September 2013 to 31 May 2016. Only ten months of overlap exist, out of the 33 months considered in this study, and the 22 considered in the previous study. Consequently, it is possible that some of the disparities discovered are due to changes in radiation belt activity. Fluctuations in space weather often cause some time periods to be more active than other.

Secondly, the standards used by *AA. Saikin et al.* [2015] to classify EMIC waves differ from those used in this study. The same minimum time period of 5 minutes was employed by both studies, and the same frequency range was adhered to. However, in this paper, the power threshold of $0.1 nT^2/Hz$ was used, while by *AA. Saikin et al.* [2015] used $0.01 nT^2/Hz$. As a result, many more EMIC waves events were flagged in the previous study than would have been had the criteria used in this paper been employed. Although one may suggest the possibility that this lower power threshold is conducive to the detection of a greater proportion of $He^+$-band waves, *AA. Saikin et al.* [2015] also found that "$He^+$-band EMIC waves average the highest wave power overall (>$0.1 nT2 /Hz$). Thus, this difference in power threshold alone does not explain the disparity in the calculated proportion of $He^+$-band waves.

The second study that shows some disagreement with my analysis was performed by *Chapmann et al.* [2016]. While its findings that helium events are most common and oxygen events are least common likewise proved consistent with the findings of this paper, the precise distributions were again slightly off. *Chapmann et al.* [2016] found 40.36% $H^+$-band waves, 57.14% $He^+$-band waves, and a mere 2.5% $O^+$-band waves.

*Chapmann et al.* [2016] also used the Van Allen probes (and hence low L-shells), and also examined a time period different from that of this study. The range of time they covered, 1 February 2015 – 23 July 2015, was rather short, and entirely covered by this study. It is likely that the small sample of data (208 events discovered in less than 6 months) examined by their study resulted in less accurate statistics than those calculated in this paper. It should also be noted that *Chapmann et al.* [2016], much like the aforementioned 2015 study, employed the power threshold of $0.01nT^2/Hz$, but, as discussed above, this is not likely to account for the greater proportion of helium events and smaller proportion of hydrogen events.

Taken together, the statistical results for $O^+$-band waves in all three studies suggest that the most probably explanation for the discrepancies is simply that EMIC wave band distributions vary with time. The 2015 study found 8.82% oxygen events, the 2016 study found 2.5%, and this study found a proportion in between those two, 5.05%. No one result is significantly different from both of the others, and consequently, none of these studies are likely to be flawed in any substantial way.

A different study, performed by *Fraser et al.,* [2010], is more supportive of the equal proportions of hydrogen and helium events found in this paper. *Fraser et al.,* [2010] examined band distribution at different phases of geomagnetic storms. The overall band distribution they observed was 52% $He^+$-band waves, and 48% $H^+$-band waves (they did not include the oxygen wave-band in their study). This is a nearly equivalent proportion, much like the distribution seen in my analysis. Furthermore, during the recovery phase of storms, they detected a considerable majority of hydrogen events, 61% $H^+$-band to only 39% $He^+$-band waves. Considering such findings, the proportions calculated in this paper are not particularly anomalous.

The examination of bandwidth and time duration distributions performed in this study has no precedent, and thus no direct comparisons can be made. *Chapmann et al.* [2016] did provide a breakdown of average bandwidth, but they did this by month (e.g., the average bandwidth of all waves observed in September is x), and did not do this for each wave-band. Oddly, their results were considerably higher than those of this paper; they found that most of the average bandwidths to be in the range 0.6 - 0.8 Hz.

They also calculated average wave duration by month, and total time durations of each band; these results are more in agreement with my findings. The average time durations tended to lie within the range 25 – 40 minutes; similarly, the most populated bin in my histogram is 20 – 40 minutes. Their calculation of total time durations found that helium wave-band events exhibited the longest observation time. While this result agrees with my analysis, it is possibly only a reflection of the fact that *Chapmann et al.* [2016] simply observed many more $He^+$-band events than $H^+$-band events, unlike me.

Overall, my results reveal that the different wave-band of EMIC waves do in fact exhibit different preferences for frequency range and time duration. Specifically, it has been found that $H^+$-band events are unique in that they, on average, exhibit a greater bandwidth and shorter time duration than the other two bands of EMIC waves. The difference in bandwidth is particularly pronounced; only a small minority of hydrogen events had a bandwidth less than or equal to 0.25

Hz (43 of 1,155), for example, while a considerable majority of helium and oxygen events occurred in this bin of the histogram above (459 of 1,176, and 81 of 125, respectively).

## 5. Conclusion and Future Work

This study provides a statistical analysis of EMIC waves, focusing particularly on distributions by frequency range, time duration, and band. 2,500 events captured by the Van Allen probes between 1 September 2013 and 31 May 2016 have been examined. A plurality of these waves (47.88%) have been found to be $He^+$-band waves. A slightly smaller proportion are $H^+$-band waves, and only a small (5.05%) minority of the detected waves are of the $O^+$ wave-band.

The most common bandwidth range of EMIC waves in general is 0.26 – 0.5 Hz. This is also true of $H^+$-band waves, but not of the other two wave-bands; Helium and Oxygen wave-band events more often have a bandwidth between 0.1 and 0.25. Thus, hydrogen wave-band events tend to have greater bandwidths than other types of EMIC waves.

The most prevalent range of time durations for EMIC waves is 21 – 40 minutes. This is also true of all three wave-bands. Distributions of the other time duration ranges were not equal for all wave-bands, however, with $H^+$-band events showing a preference for 5 – 20 minutes, and the other two wave-bands more frequently having a duration between 41 and 60 minutes. Thus, hydrogen wave-band events appear to have shorter time durations on average than other types of EMIC waves.

This study has uncovered more information regarding the preferred bands, frequency ranges, and time durations of EMIC waves. In this way, it adds to previous papers such as Yu et al. [2015] and *AA. Saikin* et al. [2015], which, while investigating various other characteristics of the waves (e.g., MLT and L-shell distribution), did not analyze distributions of bandwidth and time duration. In general, this paper increases understanding of EMIC waves and hence the workings of the Earth's magnetosphere.

Limitations of this study include the resolution used; although 64 vectors per second is quite high by today's standards, instrumentation is always improving. When more advanced detection methods are implemented, one may expect slight changes in the statistics discussed in this paper. The L-shell range of the Van Allen probes also limits the applicability of the study. While the distribution of wave-bands, for example, may be approximately 48% - 47% - 5% in L-shells 2 – 8, this may not hold true in high L-shells, such as those observed by the Cluster and THEMIS spacecraft [*Allen et al., 2015; Min et all, 2012*].

Beyond using more sensitive instrumentation and observing higher L-shells, future work could also examine additional characteristics of these 2456 EMIC wave events. Specifically, the MLT and L-shell of the waves observed in this same time period could be recorded (as this information is also included in the daily spectrograms), and a statistical analysis similar to the one above could be conducted. Plots displaying the most common MLTs and L-shell for EMIC wave occurrence in general and then by wave-band could be created. The power distribution could also be investigated, by band, time (MLT), and location (L-shell). Additionally, the normal angles and ellipticity of waves could be calculated.

The reason for the trends discovered in this paper, namely, the larger frequency range and shorter time durations of $H^+$-band waves, remains an unanswered question. This, along with the reasons for the abundance of $He^+$-band and $H^+$-band waves and the relative rarity of $O^+$-band events, may be investigated in future work. A comparison between the 33 months observed here with a different 33-month period of the mission could also be carried out.